# Experimental Investigation of Distant Cellular Interaction among Adipose Derived Stem-Cells

B. Hashemibeni[1], M Sadeghian[1], M Aliakbari[1], Z Alinasab[2], M. Akbari[2], V. Salari[3]

1- Department of Anatomical Sciences, Medical School, Isfahan University of Medical Sciences, Isfahan, Iran

2- Department of Medical Physics, Isfahan University of Medical Sciences, Isfahan, Iran

3- Department of Physics, Isfahan University of Technology, Isfahan *84156-83111*, Iran

Corresponding author:
Email: vahidsalari@cc.iut.ac.ir
Phone: 0098-3133913709
Address: Department of Physics, Isfahan University of Technology, IUT street, Isfahan, Iran
Postal Code: 8415683111

## Abstract

In addition to chemical and mechanical interactions between cells electromagnetic field produced by cells has been considered as another form of signaling for cell-cell communication. The aim of this study is evaluation of electromagnetic effects on viability of Adipose-derived stem cells (ADSCs) without co-culturing.

In this study, stem cells were isolated from human adipose tissue enzymatically and proliferated in monolayer culture. Then, $5×10^4$ adipose-derived stem cells were cultured in each well of the test plate. In the first row (4 wells), ADSCs as inducer cells were cultured in DMEM[1] with 10 ng/ml Fibroblast growth factor (FGF). In adjacent and the last rows, ADSCs were cultured without FGF (as detector cells). After the three and five days the viability of cells were evaluated. Moreover, ADSCs were cultured in the same conditions but the inducer cells were placed once in the UV-filter tube and once in the quartz tube to see whether there is electromagnetic interaction among cells.

Inducer cells caused significant cell proliferation in adjacent row cells (p-value<0.01) in the fifth day. However, using the UV-filter tube and quartz tube both reduced the effect of inducer cells on adjacent cells significantly.

As a conclusion, we could detect distant cellular interaction (DCI) among adipose derived stem cells (ADSCs), but it was not electromagnetic signaling. Our results show that ADSCs affect each other via volatile signaling as a chemical distant cellular interaction (CDCI).

## Introduction

Cells communicate with each other via many mechanisms. Most known mechanisms of cell-to-cell communication in the current literature involve chemical or electrical signaling. In contrast, our understanding of non-chemical, non-electrical and non-mechanical forms of communication is still under debate [1]. There is growing experimental evidence that cells and tissues may interact over distances even when chemically and mechanically isolated, probably via electromagnetic (EM) fields [2]. Stemming from the pioneering experiments of Gurwitsch in 1923 and 1924 [3], some researchers confirmed that cellular interactions can be mediated by EM fields [1]. There is no doubt that different EM fields can affect living cells and no question that living cells can generate EM fields, but the question is whether cells can affect each other via their EM fields while their environment is full of different types of strong EM fields? If the answer is positive, how cells can encode their very weak signals from those strong environmental signals? Basically, NCDCI experiments should be designed wisely to show a strong evidence for such mechanism. Moreover, the main source of this type of cellular EM radiation and EM receptors is poorly understood. All these together make the subject of NCDCI very controversial. Nonetheless, the subject of NCDCI is at the first stages of developments and still needs more accurate experimental verifications to survive.

---

[1] Dulbecco's modified Eagle's medium

A short summary of several experiments, reporting NCDCI among different types of cells as well as the effect of light on a single cell, is seen in Table 1. In this table, we briefly mention the cell types, inducer factors for getting response in cells, the experimental conditions in which the samples are tested, the spectrum in which cells may interact and finally the investigated parameters for footprints of NCDCI.

One of the possible candidates for NCDCI is ultraweak photon emission (UPE) (or biophotons) by living cells [8]. However, recently, the plausibility of such type of signaling among living cells is debated and criticized [9] especially in the visible range in which the intensity of UPE is so weak and looks unlikely to affect neighboring cells since under light condition the "competition between UPE and room-light" is not in favour of biophotons, depending on room light intensity there are billions of photons per biophoton. In fact, even if we assume that cells are using a special mechanism for NCDCI via UPE we still do not know what physics cells are using for that mechanism. Nevertheless, it is still not definite that UPE has no any role in the other regions of EM spectrum. The recent experiments have only revealed that there is a NCDCI effect but they don't know how it acts. But once we know better about how it acts we might develop medical applications to treat diseases in probably very simple ways. All these together motivated us to test NCDCI among new types of cells in a higher energy range (i.e. UV range).

The aim of this study was investigation of electromagnetic cellular interaction among adipose derived stem cells (ADSCs) in the UV region of EM spectrum. By definition, a stem cell is characterized by its ability to self-renew and its ability to differentiate along multiple lineage pathways. Adipose derived adult stem cells may provide an additional source of stem cells chondrogenesis, osteogenesis, and adipogenesis [12]. ADSCs can be obtained easily, with minimally invasive procedures. Therefore, it will be crucial to improve the isolation and expansion efficacy of ADSCs to investigate their possible clinical relevance [12]. Since stem cells have the ability of differentiation with higher cell proliferation, we are very interested to see if there is any distant cellular interaction (DCI) between stem cells. We designed a set of experiments to understand this mechanism clearly. If there is such communication between stem cells then we intend to find out what is the nature of this interaction? Is it chemical or non-chemical? Is there any footprint of EM signaling among stem cells? These are the questions that we are trying to find answer for them in this paper.

Table1. A summary of several experiments reporting non-chemical/non-mechanical distant cellular interactions as well as the effect of light on single cell.

| Cell Type | Inducer | Conditions | Spectrum | Parameters | Year/Ref |
|---|---|---|---|---|---|
| Onion root cells | Mitogenetic Radiation | Sample neighboring with actively dividing cells | UV (?) | Cell proliferation | 1924/[3] |
| Mouse Fibroblasts (3T3 cells) | IR pulsating Laser | Direct interaction between pulsating light and single cell | Near-IR | Cell movement to light | 1992/[4] |
| Intestinal epithelial cell line (Caco-2 cells) | $H_2O_2$ | Separated by containers at different distances | Not specified | Total protein concentration, NF.B activation and structural changes | 2007/[5] |
| *Paramecium caudatum* (Protozoa) | Another cell population | Darkness and separated with cuvettes and quartz | UV | Energy uptake, cell division rate and growth correlation | 2009/[6] |
| mouse fibroblasts (NIH3T3)/ Adult human micro vascular endothelial cells (HMVECad) | ----- | Both samples were mutually exposed, Seeded in separate polystyrene Petri dishes, and a black filter was placed between dishes | Not specified | Cell number and morphology | 2011/[7] |

## Materials and Methods

The study was accomplished at Anatomical Sciences Department, Isfahan University of Medical Sciences (IUMS). The materials and methods (M&M) is a generalized form of the M&M in the former published works [10-15].

### *a) The Cells*

Subcutaneous adipose tissue (~20 g) was obtained from 4 individuals (30-50 years age), under sterile conditions and transferred to the lab. Consent was obtained from the patients previously. After removing from the body, the adipose tissue was mechanically minced and washed with Phosphate buffered saline (PBS) (Sigma) and then it was digested with 0.075% type I collagenase (Sigma) solution at 37°C for 30 min. After inactivation of the collagenase with DMEM-LG (Sigma) and 10% fetal bovine serum (FBS) (Invitrogen), the cell solution was centrifuged at 1500 rpm for 10 min. The supernatant was removed and the resultant pellet was resuspended in culture medium contained DMEM-LG supplemented with 10% FBS, 1% penicillin and streptomycin (Gibco) and then cultured at 37°C and 5% $CO_2$ conditions. Medium was replaced every 4 days. When the cells reached 80% confluence, they were passaged with 0.05% trypsin/0.53 mM EDTA (Sigma) solution.

## b) The Experiment

In this study, $5\times10^4$ adipose-derived stem cells (ADSCc) in the third passage were cultured in DMEM-LG medium supplemented with fetal bovine serum 10%, penicillin and streptomycin 1% at each well of 24-wells test plate (Crystal-grade polystyrene, Gamma sterilized, SPL). In our experiments, four plates (P1, P2, P'2 and P3) were used to test the NCDCI effect among ADSCs (see Figure 1). In the first plate, P1, in the first row (4 wells), ADSCs as inducer cells were cultured in medium with 10 ng/ml Fibroblast growth factor (FGF) to increase cell proliferation. In the adjacent (second) and lasts (sixth) rows, ADSCs were cultured in medium without FGF (See Figures 1 and 2A). These cells were mechanically isolated from inducer cells and also chemically quarantined by walls and lids, however the lids did not entirely close the wells since otherwise the cells would suffocate. Walls of wells were transparent for the wavelengths in the range 240-750 nm (Tested at Dept. of Physics, Isfahan University of Technology (IUT)), see Figure 3.

In the second plate, P2, ADSCs were cultured in the same conditions but inducer cells were placed in a UV-filter tube (SA-Iran), which prevents EM transmittance in the range 150-400 nm (Tested at Dept. of Physics, IUT), see Figures 1, 2B and 3.

In the third plate, P'2, a similar set-up as P2 was used but by replacing the UV-filter with a quartz tube with similar size and geometry (Figures 1, 2B and 2C). The quartz tube is permeable in the UV range with wavelengths bigger than 170 nm (see Figure 3). In the fourth plate, P3, as control group, $5\times10^4$ adipose-derived stem cells were cultured without FGF in the first, second and sixth rows (see Figures 1 and 2D). For each series of experiment, every plate was placed separately in each of the three identical incubators (Lab-line, $CO_2$ 5%, 37°C) under exactly similar conditions in the same room. We could not place the all plates in a single incubator simultaneously since the cells could affect each other and perturb the results. The above series of experiments repeated twice for P2 and P'2 (i.e. eight replicates per treatment), and repeated four times for P1 and P3 (i.e. sixteen replicates per treatment).

After the three and five days the viability of cells were evaluated by MTT assay. Results of tests were investigated by analysis variances of one-way (ANOVA).

## C) Measuring Method

### MTT(3(4,5-dimethylthiazol-2-yl)-2,5-diphenyltetrazolium-bromide) assay

After the three and five days culturing for MTT assay, the medium of each well was removed, rinsed with Phosphate buffered saline (PBS) (Sigma), and replaced with 400 μl serum free medium and 40 μl MTT solution (5 mg/ml in PBS)( Sigma). Then it was incubated at 37°C, 5% $CO_2$ for 4 h, so that purple formazan crystals formed in the cells. Then the medium was discarded and added 400 μl DMSO (Sigma) to each well, and incubated in dark for 2 h. DMSO dissolved the formazan crystals and created a purple color solution. Then, 100 μl of the solution transferred to 96-well plate and absorbance of each well was read at 570 nm with ELISA reader (Hiperion MPR4). The assays were performed in triplicate.

**P1**

*Row 1*: Inducer Cells (FGF added), abbreviated (**P1-I**)
*Row 2*: Detector Cells (No FGF),
abbreviated (**P1-D2**)
*Row 6*: Detector Cells (No FGF),
abbreviated (**P1-D6**)

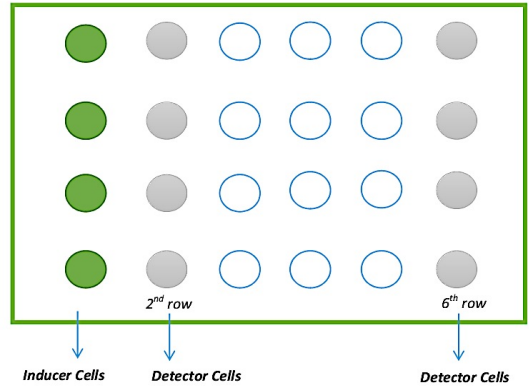

**P2**

*Row 1*: Inducer Cells (FGF added, UV-Filter used), abbreviated (**P2-I**)
*Row 2*: Detector Cells (No FGF, No Filter),
abbreviated (**P2-D2**)
*Row 6*: Detector Cells (No FGF, No Filter),
abbreviated (**P2-D6**)

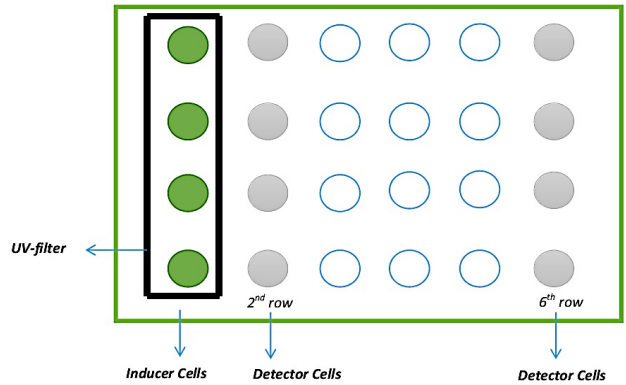

**P'2**

*Row 1*: Inducer Cells (FGF added, Quartz tube used), abbreviated (**P'2-I**)
*Row 2*: Detector Cells (No FGF, No quartz tube),
abbreviated (**P'2-D2**)
*Row 6*: Detector Cells (No FGF, No quartz tube),
abbreviated (**P'2-D6**)

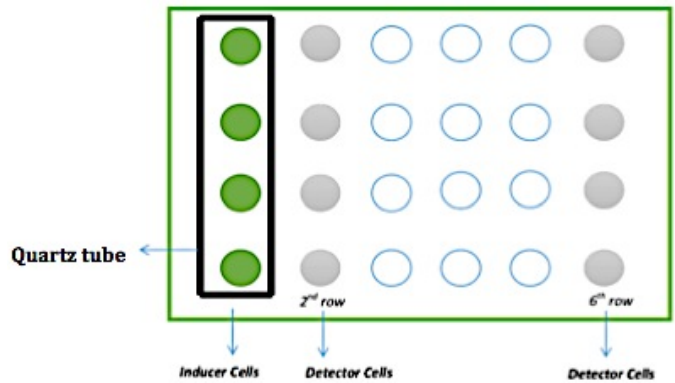

**P3**
(Control)

*Row 1*: Detector Cells (No FGF, No Filter),
abbreviated (**C1**)
*Row 2*: Detector Cells (No FGF, No Filter),
abbreviated (**(C2)**)
*Row 6*: Detector Cells (No FGF, No Filter),
abbreviated (**C6**)

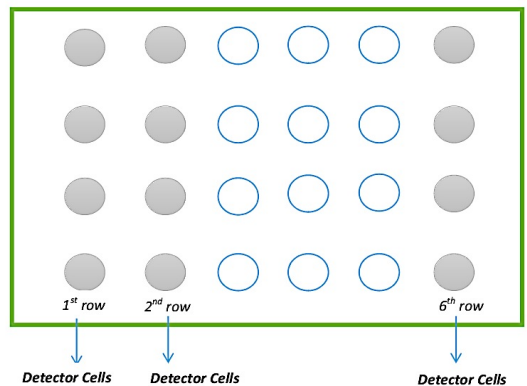

**Figure 1.** The experimental configuration of multiwell plates in which the P1 includes inducer cells in the first row, which are exposed to FGF, and detector cells are placed in the second and sixth rows respectively, P2 includes similar configuration as P1 but including UV-filter tube around inducer cells, and P'2 includes similar configuration as P2 but including quartz tube (instead UV-filter tube) around inducer cells. The control cells are placed in P3 in which there are no any FGF, no quartz and no UV-filter tubes.

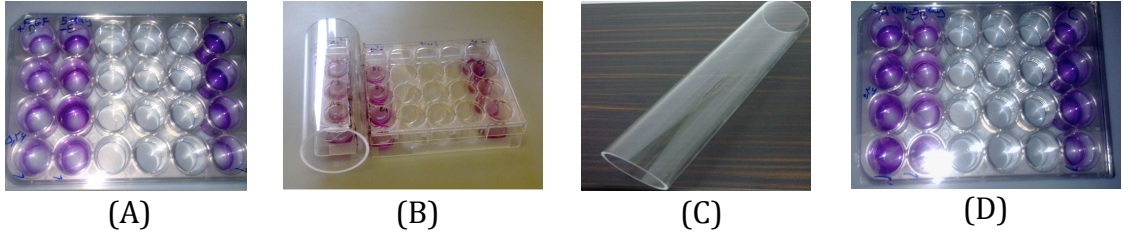

(A)  (B)  (C)  (D)

**Figure 2:** A) Plate 1 (P1), in which the adipose-derived stem cells were cultured in a medium supplemented with fibroblast growth factor (FGF) in the first row, but in the second and sixth rows the same medium without FGF was used. .B) In plate 2 (P2), the inducer cells (with FGF) in the first row were placed in the UV-filter tube. C) Quartz tube, which was used instead the UV-filter tube in a similar set-up as B (or P2), in the plate P'2. D) In plate3 (P3), as control group, in the first, second and sixth rows a medium was applied without any FGF, quartz and UV-filter.

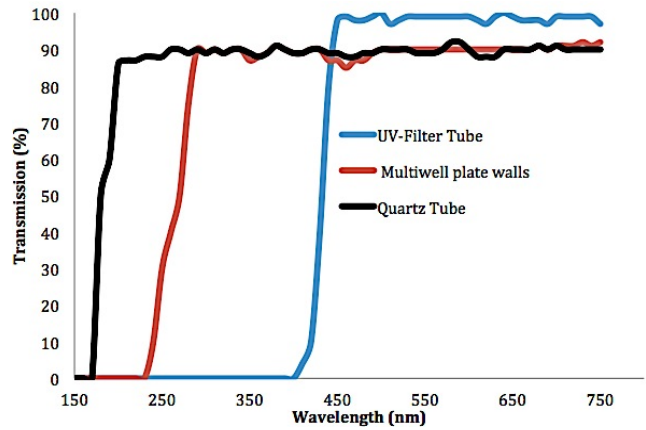

**Figure 3:** Transmission vs Wavelength Diagrams for UV-filter tube, multiwell plate walls and quartz tube.

## Results

The results of MTT method at the third and fifth days were obtained in terms of optical density (OD) after culturing. Then, we determined the viability of cells based on the OD data via below formulation,

$$viability = \frac{OD_{Test}}{OD_{Control}} \times 100 \quad (1)$$

where $OD_{Test}$ is the optical density of testing cells and $OD_{Control}$ is the optical density of the control cells. In summary, we investigate each row cells with abbreviations as follow: in the first plate (P1): (P1-I=inducer cells in the first row in P1), (P1-D2=detector cells in the second row of P1), (P1-D6=detector cells in the sixth row of P1). In the second plate (P2): (P2-I=inducer cells in the first row in P2), ((P2-D2)=detector cells in the second row of P2), ((P2-D6)=detector cells in the sixth row of P2). In the third plate (P'2): (P'2-I=inducer cells in the first row in P'2), ((P'2-D2)=detector cells in the second row of P'2), ((P'2-

D6)=detector cells in the sixth row of P'2). Control cells were in the fourth plate (P3) and the abbreviation of control, C, is used: ((C1)=control cells in the first row in P3), ((C2)=control cells in the second row of P3) and ((C6)=control cells in the sixth row of P3). The results in the fifth day are plotted in Figure 4. We obtain significant results based on the mean viability in which p-value<0.05. Our analysis indicates that there are no significant results in the third day (data are not shown), but in the fifth day several significant results appeared, which we investigate them in the following.

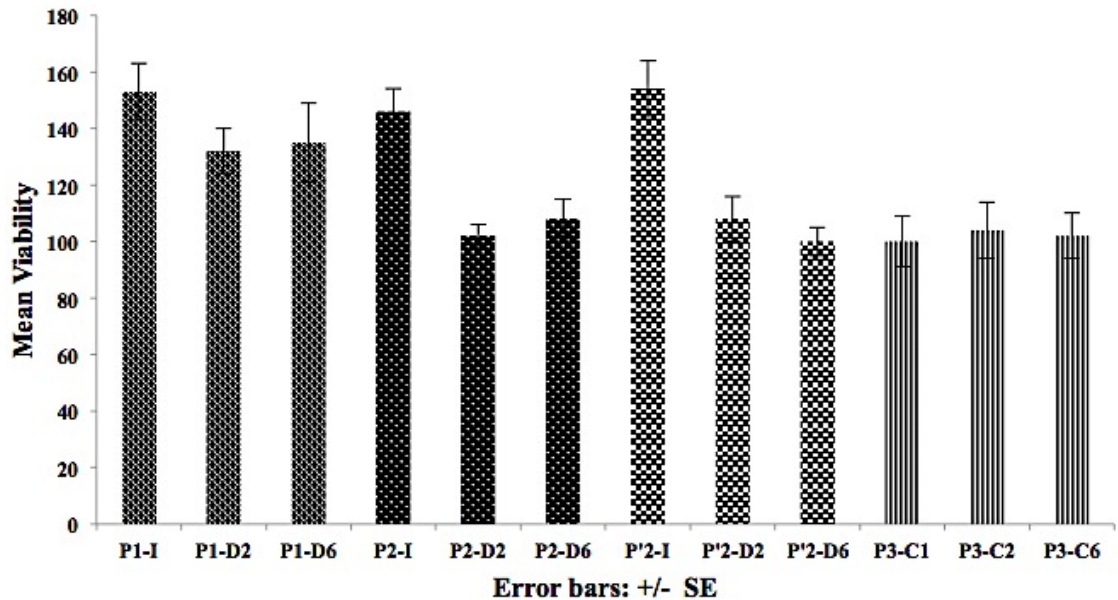

**Figure4**: The histogram of MTT results in terms of mean viability in the 5th day. Significant differences (pvalue<0.01) were observed between detector cells in P1 (P1-D2, P1-D6) and detector cells in P3 (P3-C2, P3-C6), which indicates there is a distant cellular interaction (DCI) among cells in P1. However, no significant results were observed (pvalue>0.09) among detector cells in P2 and P'2 (P2-D2, P2-D6, P'2-D2, P'2-D6) and control cells in P3. It indicates that both the UV-filter tube and quartz tube vanished the effect of DCI among inducer cells and detector cells, so this means the interaction is not electromagnetic-type.

The analysis of results indicated that there were significant results between inducer cells in P1, P2 and P'2 (P1-I, P2-I, P'2-I) and control cells (P3-C1, P3-C2, P3-C6) in P3, i.e. pvalue<0.009. This means that adding FGF to inducer cells increased cell division significantly. However, no any significant difference was observed between detector cells in P2 (P2-D2, P2-D6), detector cells in P'2 (P'2-D2, P'2-D6), and control cells in P3 (P3-C1, P3-C2, P3-C3), i.e. pvalue>0.09. However, there were significant results between detector cells in P1 (P1-D2, P1-D6) and control cells, i.e. pvalue<0.01. Moreover, there was no significant difference (pvalue=0.07) between detector cells in P1 in the second row (P1-D2) and the sixth row (P1-D6), which didn't show the effect of distance from inducer cells. These results show that there was distant cellular interaction among inducer cells and detector cells in P1. On the other side, the effects of both quartz and UV-filter tubes were similar that inhibited interaction among inducer cells and detector cells in P2 and P'2. This means that the interaction among cells in P1 was not because of electromagnetic radiation of the cells. So, the question is "what mechanism is the cause of DCI?"

## NCDCI or CDCI?

It has been discussed that cell-to-cell signaling can also be based on volatile (i.e. distant but chemical communication) which has already been demonstrated to take place between several prokaryotic as well as eukaryotic microorganisms [16] (e.g. yeast [17-20], *Escherichia coli* [21, 22], *Bacillus licheniformis* [23], *Candida albicans* [24], *Trichoderma* [25], *Serratia rubidaea* [26], *Chlamydomonas reinhardtii* [27]) and plants [28-30].

One may ask here whether the above DCI affection in our experiments is because of NCDCI or chemical distant cellular interaction (i.e. CDCI) due to volatile signaling of chemicals in the inducer cells? Indeed, the multiwell plates had lids, which inhibited the volatile signaling very much, but CDCI is still possible since the lids didn't close the top of the wells entirely, thus the volatile signaling still exists but the amount of that looks low. We could not close the top of the wells totally since otherwise the cells would suffocate. Now, the question is: which signal is more probable for DCI? NCDCI or CDCI?

## Quantitative analysis of volatile signaling

We would like to investigate the amount of vapor propagation quantitatively. The best theory to explain vapor propagation in terms of time and distance is the Fick's laws of diffusion [31]. In diffusion, mass transfer occurs via random movements at the molecular level.

*a) The Fick's first law:*

This law explains how a gas will move from a region of high concentration to a region of low concentration across a concentration gradient under the assumption of steady state. The Fick's first law is

$$J = -D\nabla C \quad (2)$$

where $J$ is the current flux, $D$ is diffusion coefficient and $\nabla C$ is the concentration gradient. In one dimension, $J_x = D\frac{\partial C}{\partial x}$ where $J_x$ is the diffusion flux per unit of area (area perpendicular to $x$), $C$ is concentration and $x$ is the distance. In a simpler form it becomes

$$J_x = D\frac{C_2 - C_1}{x_2 - x_1} \quad (3)$$

where $C_2$ is the higher concentration and $C_1$ is the lower concentration between the two points $x_2$ and $x_1$. Based on the Fick's first law, diffusion happens between two regions when there is a concentration difference between those regions. First, we consider multiwell plate without adding FGF. Since the concentrations of media in occupied wells are equal then there is no diffusion between each two occupied wells. However, after adding FGF to the wells of the inducer row the concentration difference causes diffusion of FGF from the wells of the inducer row to the adjacent rows. The amount of concentration difference is the amount of FGF in the inducer row (i.e. 10 ng/ml=$10^{-2}$ gr/m$^3$ according to M&M).

*b) The Fick's second law:*

In reality, the concentration of medium is varying with time and thus the diffusion process should be modeled according to the Fick's second law, which is in the following form in one-dimensional coordinates:

$$\frac{\partial C}{\partial t} = \frac{1}{x}\frac{\partial}{\partial x}(Dx\frac{\partial C}{\partial x}) \qquad (4)$$

where $C$ is the local concentration of the chemical, $t$ is the time, $x$ is the distance (i.e. the radius relative to the center at source), and $D$ is the diffusion coefficient.

The concentration profile is obtained from solving numerically the integral of equation (4) where a possible solution is given by:

$$C(x,t) = C_0(1 - erf(\frac{x}{2\sqrt{Dt}})) \qquad (5)$$

where $C(x, t)$ is the concentration at point $x$ and time $t$, $C_0$ is the concentration of the source, and $erf(x)$ is the error function. As we discussed earlier, $C_0$=0.01gr/m$^3$ for the primary concentration of FGF at each well of the inducer row. The diffusion coefficient of FGF at 37$^0$C is $D_{FGF}$=1.32×10$^{-4}$ cm$^2$/min(=22×10$^{-11}$m$^2$/s) [32]. We have plotted the 3D diagram of equation (5) in Figure 5 that is the concentration of FGF vapor in terms of distance and time. This diagram is for the state in which there is no barrier and walls against the diffusion of FGF, so the real values of FGF vapor concentration will be less than the values in the diagram. It is seen that during five days the FGF vapor can only diffuse maximally 3 cm distant from the source and the concentration of diffused FGF is less than 0.00005gr/m$^3$, which is a trivial value relative to the initial values in the inducer row. To be more accurate, we have determined the concentration of diffused FGF at each well of the detector rows. The results are shown in the Table 2. These results are maximum possible estimations because we ignored the geometry of barriers against volatile movements. So, considering about 10 cm distance between the first and sixth rows (see Figure 6) as well as the existence of walls and barriers against diffusion we can make sure that FGF volatile signaling cannot affect the sixth row after five days. However, our results indicate that inducer cells in P1 also affected the sixth row and there was no significant difference (pvalue=0.07) between detector cells in P1 in the second row (P1-D2) and the sixth row (P1-D6). This makes the problem more complicated, as FGF cannot diffuse to the sixth row while inducer cells affect it.

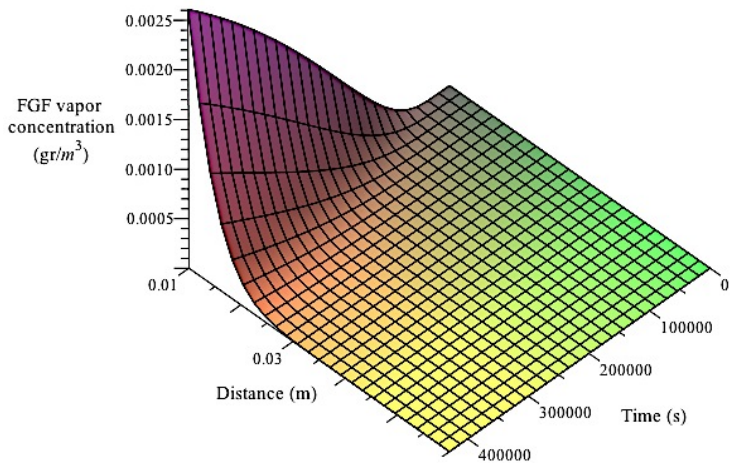

**Figure 5-** The diffused FGF vapor concentration at different distances in terms of time, based on the Fick's second law. It shows that FGF cannot diffuse more than 3cm after five days.

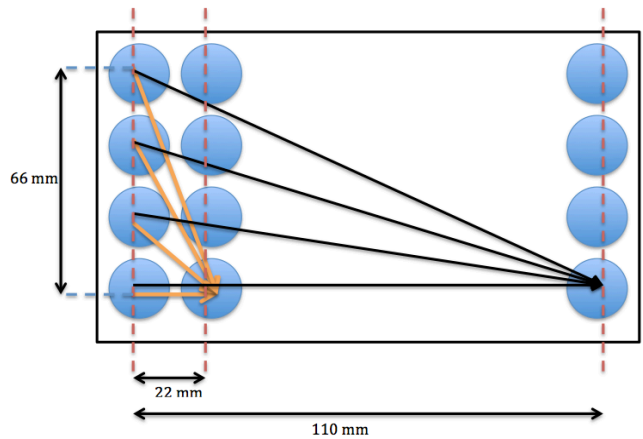

**Figure 6-** The top view of the multiwell plate structure for the distances between the wells and rows. The concentration of FGF vapor at each well of the detector rows is the summation of the diffused vapor from the all four wells in the first row (i.e. inducer row). The magnitudes of multiwell plate dimensions are shown.

Table 2. The maximum possible estimations for concentration of diffused FGF at each well of the detector rows (2nd and 6th rows) in P1 after different time intervals.

| Wells \ Time | Concentration of FGF vapor (gr/m$^3$) | | | | | | | |
|---|---|---|---|---|---|---|---|---|
| | Second Row (P1-D2) | | | | Sixth Row (P1-D6) | | | |
| | R2-1 | R2-2 | R2-3 | R2-4 | R6-1 | R6-2 | R6-3 | R6-4 |
| 10 hours | 0 | 0 | 0 | 0 | 0 | 0 | 0 | 0 |
| 1 Day | $3.4 \times 10^{-10}$ | $3.4 \times 10^{-10}$ | $3.4 \times 10^{-10}$ | $3.4 \times 10^{-10}$ | 0 | 0 | 0 | 0 |
| 2$^{nd}$ Days | $9.5 \times 10^{-7}$ | $9.5 \times 10^{-7}$ | $9.5 \times 10^{-7}$ | $9.5 \times 10^{-7}$ | 0 | 0 | 0 | 0 |
| 3$^{th}$ Days | $14 \times 10^{-6}$ | $14 \times 10^{-6}$ | $14 \times 10^{-6}$ | $14 \times 10^{-6}$ | 0 | 0 | 0 | 0 |
| 4$^{th}$ Days | $5.8 \times 10^{-5}$ | $5.9 \times 10^{-5}$ | $5.9 \times 10^{-5}$ | $5.8 \times 10^{-5}$ | 0 | 0 | 0 | 0 |
| 5$^{th}$ Days | 0.000140 | 0.000145 | 0.000145 | 0.000140 | 0 | 0 | 0 | 0 |

Indeed, we cannot determine which gas can act as signaling molecules in our experiments because our setup is not suitable for understanding this mechanism in detail. Here, we can only make sure that there is some type of volatile signaling between cells, nevertheless obtaining the real mechanism and investigation of gas candidates for DCI among ADSCs needs another experimental setup which is beyond our paper here, but possibly can be a research subject for future works.

Now, we would like to theoretically investigate the possibilities of volatile affection in our experiments. Volodyaev et al. [20] recently showed that stimulation of budding and culture growth in yeast cell cultures could be mediated by volatile carbon dioxide ($CO_2$) as a factor of DCI. When the authors separated the cultures by metal, glass and quartz glass plates, the effect disappeared, indicating the solely involvement of volatile communication in the causation of this effect [16]. $CO_2$ sensitivity of mammalian cells has been investigated in detail in [33], though in the experiments of Volodyaev et al., in the opposite to the effects of $CO_2$ sensitivity (where an increment of the concentration of $CO_2$ suppressed mitosis and stimulated cell differentiation and invasion) they have shown that $CO_2$ stimulates budding and culture growth [20]. In our experiments, the multiwell plates were placed in incubators including 5% concentration of $CO_2$ gas (i.e. similar amount as in the work of Volodyaev et al [20]). Now, we would like to discuss whether this amount of $CO_2$ is probably able to mediate DCI among ADSCs or not. The diffusion coefficient of $CO_2$ in the air is $D_{CO2}$=16 mm²/s (=16×10$^{-6}$m²/s) [34]. Expressing the concentration in unit of % and considering the primary $CO_2$ concentration 5%, we have plotted the 3D diagram of the $CO_2$ diffusion versus time and distance (see Figure 7). It is seen that $CO_2$ is highly diffusive in the air even in short timescales. We can expect that after placing the multiwell plates in the incubator the $CO_2$ gas can enter the wells quickly despite the existence of walls and barriers (see Figure 7). Thus, the role of $CO_2$ gas for mediation of DCI in our experiments looks probable as well. There can be several candidates here for volatile signaling based on the diffusion coefficients of gas molecules. Normally, the typical values for diffusion coefficients of gases in the air are in the order of ≈10$^{-5}$m²/s [35]. A list of diffusion coefficients for different gases at 300K is seen in Table 3. Thus, there would be several candidates for volatile signaling between cells as they have high diffusivity in environment. On the other side, the biological agents like FGF have lower diffusivity relative to the gas molecules and therefore they will probably not affect neighboring cells directly as they are heavier molecules and not be diffused easily at room temperature. So, some gas molecules (e.g. in Table 3) may be chemical messengers from inducer cells to detector cells.

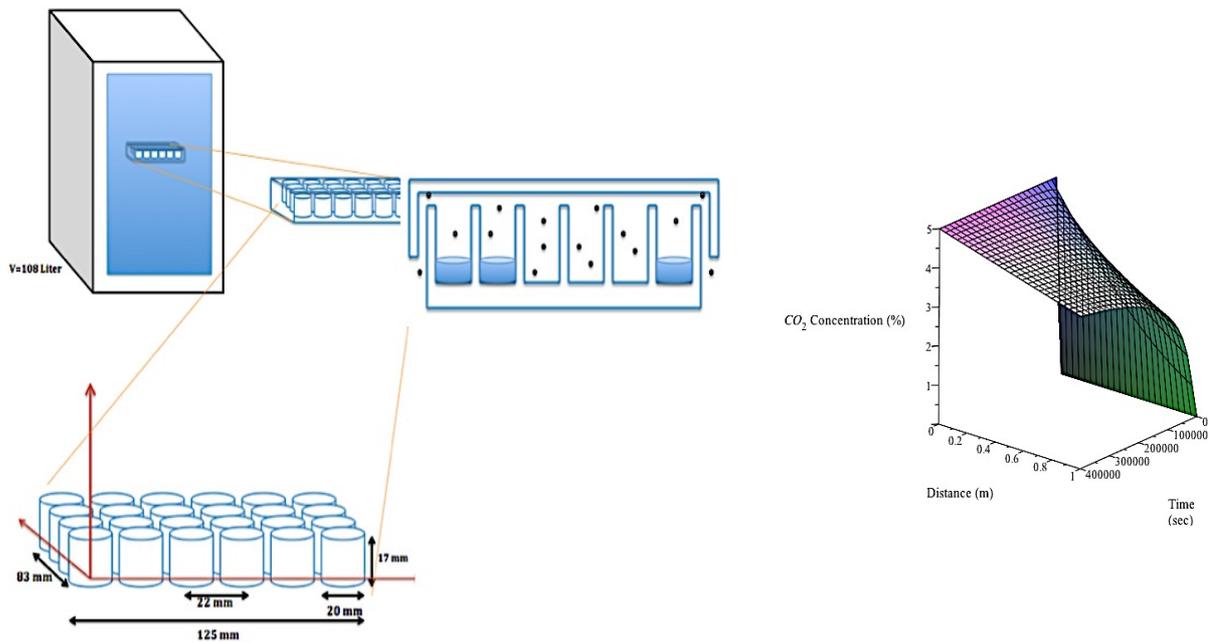

**Figure 7-** Left) The multiwell plates are placed in an incubator with 5% $CO_2$ concentration. $CO_2$ is highly diffusive and after placing the multiwell plates in the incubator the $CO_2$ gas can enter the wells quickly despite the existence of walls and barriers. Right) The 3D diagram of $CO_2$ diffusion versus time and distance. The primary concentration is considered 5%. It indicates that $CO_2$ molecules can diffuse to long distances in short times with high concentration.

Table3. Diffusion coefficient of different gas molecules at T=300K [35].

| Gas molecule | Environment | Diffusion Coefficient (m²/s) |
|:---:|:---:|:---:|
| $H_2O$ | air | $24 \times 10^{-6}$ |
| $CO_2$ | air | $14 \times 10^{-6}$ |
| CO | air | $19 \times 10^{-6}$ |
| $H_2$ | air | $78 \times 10^{-6}$ |
| $H_2$ | $O_2$ | $70 \times 10^{-6}$ |
| $H_2$ | $CO_2$ | $55 \times 10^{-6}$ |
| $O_2$ | air | $19 \times 10^{-6}$ |
| He | air | $71 \times 10^{-6}$ |
| $SO_2$ | air | $13 \times 10^{-6}$ |

One may question here whether there may be a problem regarding the two sides of the UV-filter tube and quartz tube which were open for P2 and P'2 plates in our experiments since the volatile signaling still could affect detector cells? (see figure 8). We discuss now that the effect of volatile signaling in this case is considerably low and trivial.

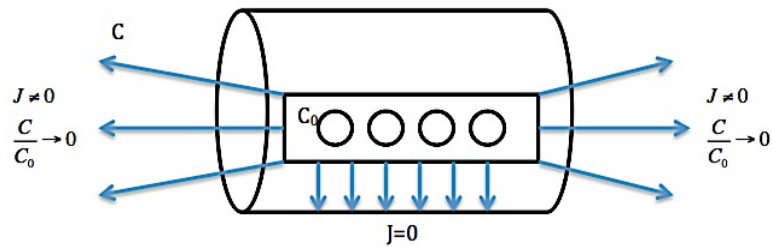

**Figure 8-** The diffusion flux (J) cannot reach to the multiwell plate considerably since the wall of the cylinder is a strong barrier against volatile. Two open sides of the cylinder is a scape way for signaling molecules, however because of the low concentration of volatile it acts like an ideal gas, so the density of volatile in incubator becomes trivial in comparison with the volatile in the multiwell plate.

We can estimate the concentration of volatile molecules (C) in terms of the number of molecules (N) per volume (V) in which molecules are moving (i.e. C=N/V). For the sake of simplicity, we consider that the number of volatile signaling molecules (produced or affected by inducer cells) is constant in a specific time interval. So, if we assume that the primary concentration of volatile is $C_0$ in the volume of wells of the inducer row (i.e. $V_0$) then the secondary concentration of volatile (C') outside of the tube (V') will be C'=$C_0$($V_0$/V'). Considering the volume of four wells in the inducer row for $V_0$ (see Figures 9 and 13) we obtain $V_0 \approx 26 \times 10^{-6} m^3$, and the incubator has the volume V'=108lit=0.18$m^3$. Thus, we obtain C'≈0.0001$C_0$. This concentration of molecules is unimportant relative to the first concentration in the wells. Moreover, the molecules outside the tube with such low concentration intend to be propagated in the big volume of incubator (based on the second law of thermodynamics) instead going directly into the wells of the detector cells. Thus, the probability of volatile signaling in the case of the presence of tubes is trivial and much less than the case without tubes.

## Conclusion:

In this paper, we have investigated distant cellular interaction among adipose derived stem cells experimentally. We have used Fibroblast growth factor (FGF) to increase the rate of cell division in inducer cells. Our results indicated that inducers cells could affect distant neighboring cells to increase cell division. To understand the nature of this signaling we isolated inducer cells once by UV-filter tube and once by quartz tube (i.e. permeable to UV) to see whether it is electromagnetic (EM) signaling or not. *No significant difference was observed*. Consequently the hypothesis of electromagnetic distant cellular interaction, or non-chemical distant cellular interaction (NCDCI), was not confirmed by our experiments. Our results besides our theoretical quantitative analysis indicate that distant cellular interaction (DCI) among ADSCs is chemical due to volatile signaling. However, obtaining the real candidate for volatile molecules as well as detailed mechanism of signaling need another experimental investigation that would be a potential prospect of future research in this context.

**Acknowledgements:**

This study was supported by Isfahan University of medical Sciences. The authors would like to express their gratitude to Isfahan University of Medical Sciences Research deputy for collaboration.